# Nanoconstriction Microscopy of the Giant Magnetoresistance in Cobalt/Copper Spin Valves

S.J.C.H. Theeuwen, J. Caro, K.P. Wellock, and S. Radelaar

*Delft Institute of Microelectronics and Submicron Technology (DIMES), Delft University of Technology, Lorentzweg 1, 2628 CJ Delft, The Netherlands*

C.H. Marrows and B.J. Hickey

*Department of Physics and Astronomy, University of Leeds, Leeds LS2 9JT, United Kingdom*

V.I. Kozub

*A.F. Ioffe Physico-Technical Institute, St.-Petersburg 194021, Russian Federation*

We use nanometer-sized point contacts to a Co/Cu spin valve to study the giant magnetoresistance (GMR) of only a few Co domains. The measured data show strong device-to-device differences of the GMR curve, which we attribute to the absence of averaging over many domains. The GMR ratio decreases with increasing bias current. For one particular device, this is accompanied by the development of two distinct GMR plateaus, the plateau level depending on bias polarity and sweep direction of the magnetic field. We attribute the observed behavior to current-induced changes of the magnetization, involving spin transfer due to incoherent emission of magnons and self-field effects.

PACS: 73.40.Jn, 73.50.Jt, 75.70.Kw, 75.70.Pa,

Currently, applications of the giant magnetoresistance (GMR) are realized, particularly in read heads for harddisks[1] and magnetoresistive random access memories (MRAMs).[2] These applications are based on the spin valve. This is a trilayer structure, with a normal metal (NM) layer sandwiched between two ferromagnetic metal (FM) layers.[3] The spin-valve resistance depends strongly on the magnetic-field-induced relative orientation of the magnetizations of the FM layers. This GMR arises predominantly from spin-dependent scattering of the electrons at the interfaces between the NM and FM layers.[1]

So far, most spin-valves studies were performed on large structures, up to millimeters wide. Since the domain size of an FM film ranges from hundreds of nanometers to several micrometers, the GMR arises from averaging over many domains. Currently, emphasis shifts to spin valves patterned into micrometer and sub-micrometer structures, entering the regime of only a few domains. This has led, for example, to the direct observation[4] of the correlation of domain structure and GMR and to very small memory elements for MRAMs.[2]

A very attractive way to enter the few-domain regime is to use a point contact to a macroscopic spin valve. Such a contact is a nanoconstriction between a metallic electrode and the spin valve. When a current flows through the device, the current density peaks strongly in the constriction (diameter $d$ =10-50 nm). As a result, in a resistance measurement a hemi-spherical region of the spin valve is probed, in size comparable to the constriction diameter (Fig. 1a). In this region only two domains are expected to contribute to the GMR, one from each FM layer. Due to the minute layer thickness (typically several nanometers), the current flows almost perpendicularly through the layers (CPP geometry), as in the multilayer pillars of Gijs *et al.*[5] and in the spin valves sandwiched between superconducting electrodes of Pratt *et al.*.[6] In this Letter we report GMR measurements on Co/Cu spin-valve point contacts. These devices work as a microscope, revealing effects of only a few domains at the constriction. Further, the point-contact geometry is very suitable to search for effects of electron-spin transfer to the FM layers,[7] either as predicted by Slonczewski[8] or by Kozub *et al.*.[9]

We fabricate the spin valves on silicon membranes.[10] On the lower membrane face a Cu(2 nm)/Co(3 nm)/Cu(3.5 nm)/Co(300 nm)/Cu(300 nm) structure is sputtered (Fig. 1a). The 2 nm Cu layer is a buffer and the 300 nm Cu cap shunts any current-in-plane (CIP) series GMR of the trilayer film.[11] The device is completed by etching a nanohole in the membrane and deposition of Cu on the upper face, to fill the hole and form the counter electrode. The different Co thicknesses cause magnetization reversal of the layers at different fields. We make identical layered films (but with 3 nm cap) for CIP control measurements. The point-contact resistance is measured in a four-wire geometry. Resistances are in the range 0.3-8 $\Omega$.

These spin valves exhibit a clear GMR. We observe substantial device-to-device differences in the GMR curve, attaining a maximum GMR ratio of 2% (ratio=$\Delta R / R_{sat}$; $\Delta R = R_0 - R_{sat}$, $R_0$=maximum resistance, $R_{sat}$=saturated resistance). Step-like transitions in the GMR often occur. These observations agree with the absence of averaging, as expected when only a few domains are probed. In general, the GMR ratio decreases with increasing bias current. Contrary to the point contacts, the CIP films do not exhibit a GMR. This indicates that 300 nm Co is thick enough to completely shunt the CIP GMR and that consequently the point-contact GMR arises from the constriction region, where the current must traverse the Co/Cu interfaces. The CIP films do show the anisotropic magnetoresistance (AMR).[12] The AMR of the point-contacts is negligible (AMR<0.1%). Its dependence on the field orientation correlates in an AMR fashion with the average current direction through the constriction (which is along the constriction axis), from which it thus must originate. The point-contact spectra $d^2I/dV^2(V)$ of the devices, *i.e.* the voltage-dependence of the second derivative of the $I-V$ curve, are featureless. This is unlike spectra of ballistic contacts,[13] which show phonon peaks, and unlike spectra of multilayer contacts of Tsoi *et al.*,[14] which show spin-wave excitations. We conclude that electron transport in our contacts is diffusive or intermediate between ballistic and diffusive.

This Letter concentrates on one specific contact which nicely illustrates the absence of averaging over the properties of many domains. This contact has $R = 0.38\,\Omega$, which corresponds to a diameter of 50 nm,[15] and a 1.2% GMR ratio at 4.2 K. This ratio is dominated by the contribution of the thick Co layer to $R_{sat}$. After correction an upper bound of the intrinsic ratio is estimated at 5%. Fig. 1b shows the GMR curve for the in-plane field geometry, measured at $T = 4.2\ K$ and with low level ac excitation ($I_{ac} = 100\ \mu A$). This curve is much narrower than the perpendicular-field curve (not shown), in agreement with the shape anisotropy of the layers. Coming from high field, the in-plane curve shows a gradual increase before $H = 0$, a hysteretic maximum close to $H = 0$ and a steep decrease to saturation at $\mu_0 |H| = 6\ mT$. This differs from the GMR of a *multi*-domain area of a spin valve, which yields a maximum after $H = 0$ due to hysteresis of the *multi*-domain system. The curve of Fig. 1b reflects specific magnetization properties of just a few or even two domains. The curve suggests a gradual change of the magnetization of one domain to a preferential direction when approaching $H = 0$ and an abrupt switching of the magnetization to a completely aligned state at $\mu_0 |H| = 6\ mT$, involving the other domain. The GMR in Fig. 1b occur at much lower field than those of Myers *et al.* for a very similar device,[7] which however is fabricated with the pre-etched nanohole method. We argue that this method, contrary to ours, for hole sizes comparable to or larger than a layer thickness, leads to strong topography at the hole after deposition of the Cu electrode. Thus, the layers are distorted, affecting the field scale.

At elevated bias currents the picture becomes more rich, as illustrated in Fig. 1c for $I_{bias} = +10\ mA$ (positive current: electrons flow from the thin Co layer to the thick one). Two pronounced resistance plateaus develop, their height depending on field-sweep direction and current polarity. The upper plateau occurs when coming from positive $H$ and biasing with positive $I_{bias}$ or from negative $H$ and applying negative $I_{bias}$. The lower plateau appears for a reversed sweep direction or opposite current polarity. Thus, one can choose the relative magnetization direction of the domains with the magnetic history and the current polarity. The gap between the upper and lower plateau first opens and then closes when the bias current is increased from 100 μA to 100 mA, reducing the GMR ratio by a factor of six. This behaviour is plotted in Fig. 2, where $R_0$ is the plateau level. The plotted bias range corresponds to current densities up to $5 \times 10^9\ A/cm^2$, *i.e.* one hundred times higher than values in GMR read heads.[3] When converted to the GMR ratio, the lower branch of $R_0$ in Fig. 2 reaches its final level of 0.2% at 20 mA, while the upper branch decreases approximately linearly across the whole bias range. The plateaus are stable for many hours. We never observed a spontaneous transition between them. Further, the side of the gradual change of the GMR peak broadens with increasing current, while the steep side is unaffected. In Fig. 3 this is illustrated for one biasing and sweeping combination.

Joule heating might be responsible for the above behaviour, in particular the decrease of the GMR peak with increasing current and its broadening. To check this, we measure the temperature dependence (1.5-293 K) of the zero-bias ($I_{ac} = 100 \: \mu A$) GMR curve, yielding a constant GMR ratio up to 100 K and a decrease to 0.6% at 293 K (inset Fig. 3). Further, with increasing temperature the gradual side of the GMR curve is hardly affected, while the steep side shifts towards $H = 0$. This is contrary to the bias-dependence, so that Joule heating is clearly excluded. The zero-field resistance $R_0(T = 4.2K, I_{bias})$ decreases in the range 0-50 mA, whereafter it increases (Fig. 2). Again, this differs from the measured $R_0(T, I_{ac} = 100 \: \mu A)$, which increases monotonously. The measured bias-dependence $R_{sat}(T = 4.2K, I_{bias})$ also shows a monotonic increase. We attribute the latter increase to extra scattering due to phonon and magnon creation enabled by a bias-induced non-equilibrium electron-distribution function. The resulting resistance increase also contributes to $R_0(T = 4.2K, I_{bias})$. However, we argue that below 50 mA $R_0(T = 4.2K, I_{bias})$ is dominated by another effect, *viz.* current-induced magnetization changes. These arise from the self-field (*i.e.* the field generated by the current according to Ampère's law) or from spin transfer to the FM layers. Spin transfer leads to coherent rotation of the magnetization[8], resulting in switching of the resistance,[7] or to incoherent emission of non-equilibrium magnons,[9] as discussed below.

We estimate the self-field by approximating the point contact by a hyperboloid of revolution.[16] For $I = 5 \: mA$ the maximum of $\mu_0 H_{self}$ in the thin(thick) Co layer is 24(12) mT. This is comparable to the field scale of the zero-bias GMR. Thus, the self-field should affect the GMR for most of the bias range. For example, it should broaden the GMR peak. Broadening is indeed observed for the gradual side of the peak, but the steep side is unaffected. Apparently, other and stronger driving forces dominate certain features of the GMR, for example a uniaxial anisotropy induced by fabrication-related effects. This, combined with the self-field, which depending on the bias polarity can guide the magnetization in either direction of the anisotropy axis, can give rise to plateaus. With increasing bias the self-field will tend to impose a vortex character upon the magnetization of both Co layers, eventually leading to almost congruent vortices in the Co layers and consequently a reduction of the GMR, which is what we observe.

Spin transfer influences the thin Co layer much more than the thick Co layer, since the latter is strongly exchange coupled to its surroundings.[8] According to the coherent rotation model,[8] the magnetization of the thin layer and thus the MR should switch abruptly at a threshold current. Our contacts do not show switching, apparently excluding this mechanism. Rather, the GMR evolves gradually with bias. This can be explained[9] by incoherent emission of non-equilibrium magnons, giving rise to a magnon "hot spot" in the thin Co layer, characterized by an effective magnon temperature $T_m$. Since the probability of electron-magnon creation (a spin-flip process) is proportional to the density of final electron states, which in turn is spin-dependent, $T_m$ depends on the current polarity and the relative orientation of the magnetizations of the layers, angles close to 0ϒ or 180ϒ favouring high $T_m$. Further, $T_m$ is proportional to the bias. Deviations of the magnetization of the thin layer from a preferential direction are assumed to be thermally activated by $T_m$. Apart from a gradual decrease of the GMR with current, this model also gives a current-direction dependent GMR plateau. However, the plateau level does not depend on the sweep direction, contrary to the measurements. This indicates that the model in its present form does not catch all magnetic ingredients of the spin valve. In addition, a complete treatment should include the interplay of spin transfer and the self-field effect. Similar to the Slonczewski's spin transfer, also the spin-transfer effect put forward here can be used to set the bits of an MRAM, but with the additional advantage of write capability for parallel and antiparallel orientations of the magnetizations.

In conclusion, using Co/Cu spin-valve point contacts, we have studied the CPP GMR of only a few domains. Due to absence of averaging over many domains, the GMR curve shows strong device-to-device differences, the maximum GMR ratio being 2%. For one device we studied in detail the dependence of the GMR curve on the bias current up to 100 mA and on temperature up to $293\,K$. The GMR ratio decreases with increasing bias, while two distinct GMR plateaus develop around $H = 0$. We attribute this behavior to current-induced changes of the magnetization, involving spin transfer due to incoherent emission of magnons and self-field effects, while we exclude Joule heating.

We gratefully acknowledge G.E.W. Bauer, R.P. van Gorkom and T.M. Klapwijk for stimulating discussions and N.N. Gribov for expert support in preparing Si membranes.

inhomogeneous) self-field $\mu_0 H_{self} = F(\eta,\xi)(\mu_0 I_{bias}/2\pi a)$. The maximum value of the spatial function $F(\eta,\xi)$ in the thin(thick) Co layers at the constriction is approximately 0.6(0.3).

**Figure captions**

Fig. 1. Schematic cross-section of a spin-valve point contact (a) and GMR of an 0.38 Ω contact, measured at $T = 4.2\,K$ with an in-plane field, for $I_{ac} = 100\,\mu A$ (b) and $I_{bias} = 10\,mA$ (c).

Fig. 2. Dependences of $R_0$ and $R_{sat}$ on bias current, for the combinations of sweep direction and current polarity, for the in-plane field geometry and $T = 4.2\,K$. The opening and closing of a gap between plateaus can be seen. Lines guide the eye.

Fig. 3. GMR for different positive currents. $T = 4.2\,K$. The arrow indicates the sweep direction. For high currents the GMR is reduced and broadened on one side. The inset shows the temperature-dependence of the GMR ratio.

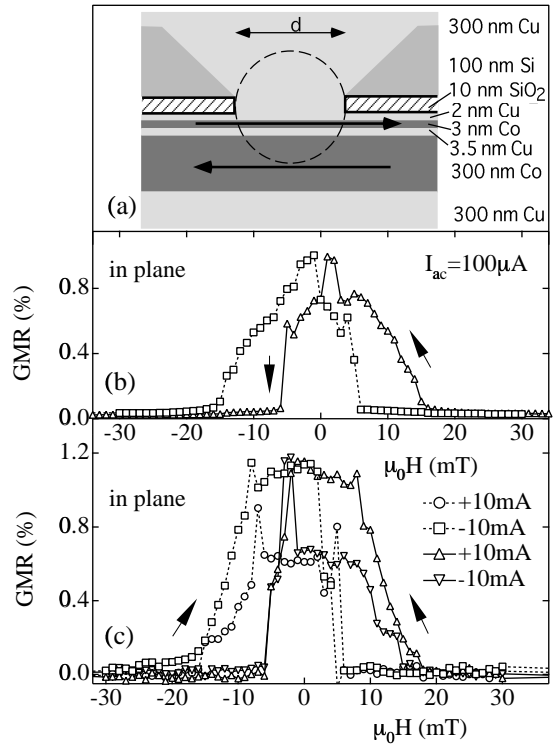

Fig. 1

Theeuwen et al.

APL

```
```

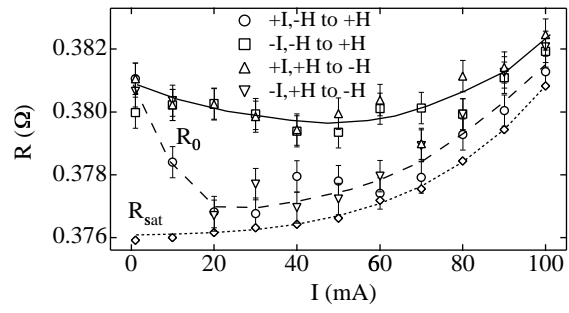

Fig. 2

Theeuwen et al.

APL

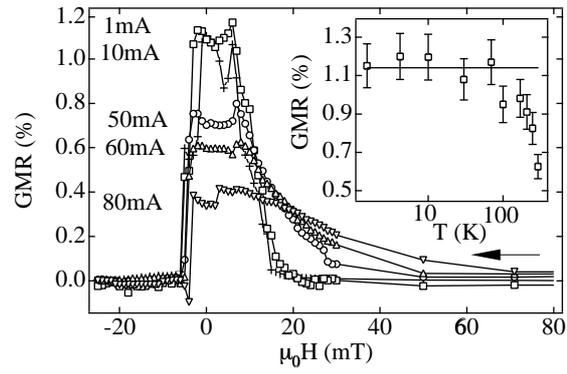

Fig. 3

Theeuwen et al.

APL